\documentclass[aps,prl,twocolumn,showpacs,superscriptaddress,groupedaddress]
{revtex4-2}  
\usepackage{graphicx}  
\usepackage{dcolumn}   
\usepackage{bm}        
\usepackage{amssymb}   
\usepackage{amsmath}   

\usepackage{verbatim}  

\usepackage{xcolor}

\begin{document}
\title{Thermal drag effect in quantum Hall circuits}

\author{Edvin G. Idrisov}
\affiliation{Department of Physics and Materials Science, University of Luxembourg, Luxembourg}

\author{Ivan P. Levkivskyi}
\affiliation{ Dropbox Ireland, One Park Place, Hatch Street Upper, Dublin, Ireland}


\author{Eugene V. Sukhorukov}
\affiliation{D\'epartement de Physique Th\'eorique, Universit\'e de Gen\`eve, CH-1211 Gen\`eve 4, Switzerland}

\date{\today}




\begin{abstract}
We study the thermal drag between two mesoscopic quantum Hall (QH) circuits. Each circuit consists of Ohmic contact perfectly coupled to quantum Hall edge states. The drag is caused by strong capacitive coupling between Ohmic contacts. The non-equilibirum conditions and the electron-electron interaction are taken into account by using the non-equilibrium bosonization technique. The thermal drag current in the passive circuit, the noise power of the corresponding heat current, and the Fano factor are calculated and analyzed.
\end{abstract}

\maketitle


The mutual electron drag effect is the phenomenon that arises in a system of  two isolated (in the absence of the transfer of charge) quantum circuits coupled via the long-range Coulomb interaction~\cite{Narozhny}. Typically, one is interested in the so-called Coulomb drag effect, where the charge current in the ``active'' part of the circuit causes the  charge drag current  in  the ``passive'' part of the circuit~\cite{Levchenko}, and normally, this effect is perturbative and weak.
The Coulomb drag effect in various systems was thoroughly studied both experimentally and theoretically~\cite{Narozhny}. It was further shown that the charge drag current can also be mediated by the combined electron-phonon, electron-photon, and electron-ion interactions~\cite{Raichev,Strait,Gurevich}. All these effect have to be differentiated with the  thermal drag effect, which can also be cause by the long-range Coulomb interaction~\cite{Joulain,Bhandari,Abdallah,Berdanier}. In this case  the flow of heat current in the active  circuit results in the heat current in the passive circuit. Of particular interest is the thermal drag effect due to the temperature difference between active and passive circuits. The effect of thermal drag was studied in the context of quantum refrigerators and engines, heat diodes and heat  pumps~\cite{Venturelli,Zhang,Koski,Whitney,Ruokola,Moskalets}. 

\begin{figure}
\includegraphics[width=0.6\columnwidth]{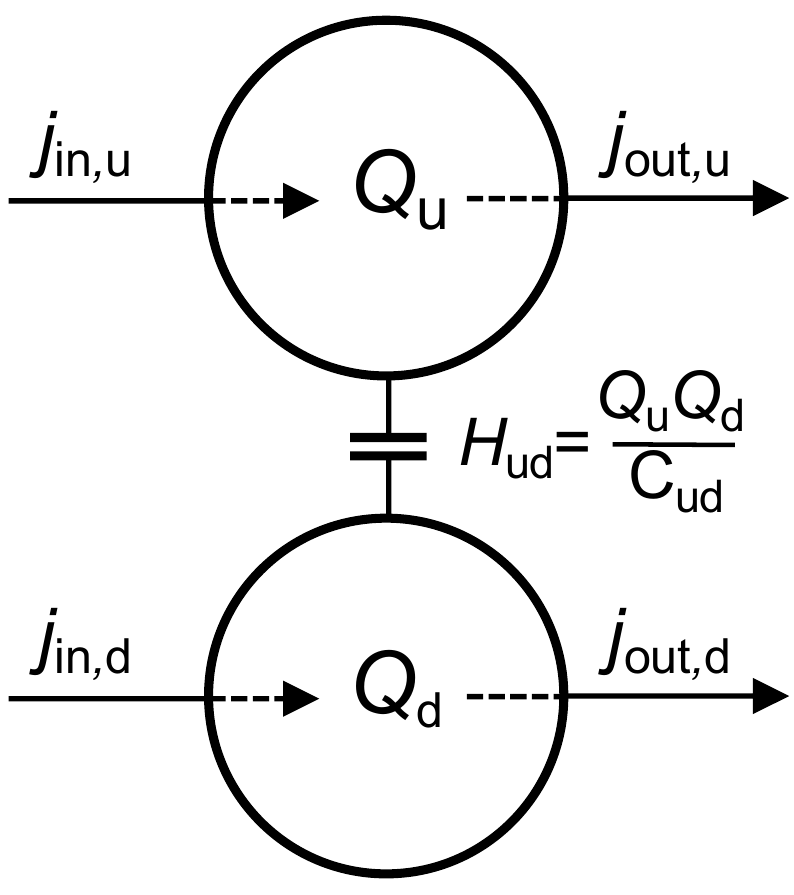}
\caption{\label{fig:one} Schematic of the possible experimental setup. The systems consists of two identical quantum circuits: upper (active) and lower (passive) parts. Each part includes an Ohmic contact (metallic granula with capacitance $C$ and negligible level spacing) which is perfectly coupled to chiral integer QH channel (the channel enters an Ohmic contact and leaves it without electron backscattering). The incoming channels are kept at different temperatures $T_{\rm in, u}-T_{\rm in, d}>0$. The capacitive coupling between Ohmic contacts, $Q_{\rm u} Q_{\rm d}/C_{\rm ud}$, results in thermal drag effect: The upper mesoscopic circuit induces extra heat current in the lower circuit in the outgoing channel.}
\end{figure}

The recent progress in the  fabrication of hybrid systems that are based on chiral  QH edge states made it possible to do accurate mesoscopic experiments with strongly interacting electron systems~\cite{Zubair}. A particular interesting example of such system is a QH edge state perfectly coupled to an Ohmic contact, a mesoscale  metallic granula with negligible level spacing and finite charging energy comparable to the  temperature~\cite{Cavanna}. It is worth mentioning that such an  Ohmic contact immediately became a key element of new experimental and theoretical studies~\cite{Zubair}, the remarkable examples being the suppression of charge quantization caused by quantum and thermal fluctuations~\cite{Jezouin2016,IdrisovThermalDecay}, the heat Coulomb blockade effect~\cite{Furusaki,ArturEquil,Sivre2018,Sivre2019}, the interaction induced recovery of the phase coherence~\cite{Clerk,IdrisovDephasing,Pierre0}, the charge Kondo effect~\cite{Iftikhar2015,Iftikhar2018}, the quantization of the anyonic heat flow~\cite{Heiblum1}, and the observation of the half-integer thermal Hall conductance~\cite{Heiblum2,Mross}.


In this Letter, we propose a mesoscopic QH system with embedded Ohmic contacts,  building up strong Coulomb interactions,  for studying thermal drag effect (see Fig.~\ref{fig:one}). The system consists of the active (upper) and passive (lower) quantum circuits. Each circuit includes an Ohmic contact perfectly coupled to a chiral QH edge state: The edge state enters an Ohmic contact and leaves it without electron backscattering. The thermal drag effect is caused by the capacitive coupling between Ohmic contacts of the active and passive circuits. The temperature difference $T_{\rm in, u}-T_{\rm in, d} > 0$ is applied between circuits which causes the temperature imbalance between contacts and results in the thermal heat flow from the upper to the lower part. We assume the full thermalization of the electron systems inside Ohmic contacts. In our study we focus on the thermal drag current, its zero-frequency noise power, and the associated Fano factor. 
   
\begin{figure}
	\includegraphics[width=0.7\columnwidth]{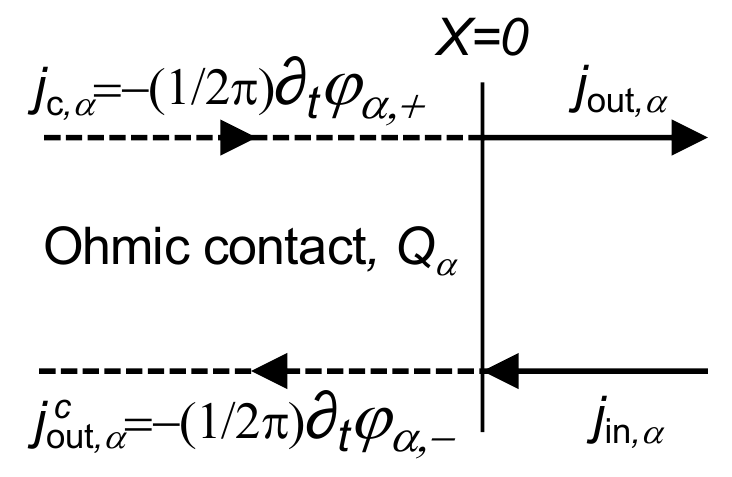}
	\caption{\label{fig:two} The model of an Ohmic contact with the capacitive interaction is illustrated, see Eq.~(\ref{Hamiltonian of each quantum circuit}). The charge current $j_{\rm in,\alpha}(t)$ arriving at the Ohmic contact enters it at the interface at $x=0$ and continues as a neutral current $j^{\rm c}_{\rm out,\alpha}(t)$. The neutral current $j^{\rm c}_{\alpha}(t)$ arriving at the interface from inside the Ohmic contact continues as a charge current $j_{\rm out,\alpha}(t)$.  In order to account for the negligible level spacing in the Ohmic contact, we extend edge states inside it to minus infinity and introduce the small regularization parameter $\epsilon$ in Eq.~(\ref{Hamiltonian of each quantum circuit}).}
\end{figure}

\textit{Model and theoretical approach}.
We use the low-energy effective field theory~\cite{Wen,Giamarchi} to describe QH edge states strongly coupled to the Ohmic contacts.
The Ohmic contact is modeled by extending an edge state inside the metallic granula and splitting it in two uncorrelated channels~\cite{Furusaki,ArturEquil}. The Hamiltonian of each quantum circuit ($\alpha=\rm u,d$) contains two terms (see Fig.~\ref{fig:two})
\begin{equation}
\label{Hamiltonian of each quantum circuit}
\begin{split}
& H_{\alpha}=\frac{v_F}{4\pi}\sum_{\sigma =\pm} \int dx (\partial_x \phi_{\alpha \sigma})^2+Q^2_{\alpha}/2C, \\
& Q_{\alpha}=\int_{-\infty}^0 dx e^{\epsilon x/v_F}\left[\rho_{\alpha +}(x)+\rho_{\alpha -}(x)\right], 
\end{split}
\end{equation}
where $v_F$ is Fermi velocity, $C$ is capacitance of each Ohmic contact, $Q_{\alpha}$ is an operator of the total charge accumulated at each Ohmic contact, and $\epsilon$ is a small regularization parameter. Here the first term accounts for the dynamics of incoming and outgoing edge channels, and the second term describes the charging energy of an Ohmic contact of a finite size.  The set of scalar fields $\phi_{\alpha \sigma}(x,t)$, where $\alpha= \rm u,d$ and $\sigma=\pm$,  are introduced in Eq.~(\ref{Hamiltonian of each quantum circuit}) to describe the low energy physics. The operators of edge charge densities and currents for incoming, $\sigma=-$, and outgoing, $\sigma=+$, states are given by $\rho_{\alpha \sigma}=(1/2\pi)\partial_x \phi_{\alpha \sigma}$ and $j_{\alpha \sigma}=-(1/2\pi) \partial_t \phi_{\alpha \sigma}$. The bosonic fields satisfy the standard canonical commutation relations  
\begin{equation}
\label{Canonical commutation relation for bosonic fields}
[\partial_x \phi_{\alpha \sigma}(x,t), \phi_{\alpha^{\prime} \sigma^{\prime}}(x^{\prime},t)]=2\pi i \sigma \delta_{\alpha \alpha^{\prime}}\delta_{\sigma \sigma^{\prime}}\delta(x-x^{\prime}),
\end{equation}
where $\delta_{\alpha \alpha^{\prime}}$ is the Kronecker delta and $\delta (x)$ is the Dirac delta function. It is worth mentioning that the presented above model of an Ohmic contact was successfully used to explained the experiments on the charge quantization, heat Coulomb blockade and heat quantization of anyonic flow~\cite{Jezouin2016,Sivre2018,Sivre2019,Heiblum1,Heiblum2,Mross}.

The total Hamiltonian of the system includes three terms (see Fig.~\ref{fig:one}) 
\begin{equation}
\label{Total Hamiltonian}	
H=H_{\rm u}+H_{\rm d}+H_{\rm ud},
\end{equation}
where the Hamiltonians of the active and passive circuit, $H_{\alpha}$, are given in Eq.~(\ref{Hamiltonian of each quantum circuit}), while capacitive coupling between Ohmic contacts has the form
\begin{equation}
\label{Capacitive coupling between Ohmic contacts}
H_{\rm ud}=Q_{\rm u} Q_{\rm d} /C_{\rm ud}, 
\end{equation}
where $1/C_{\rm ud}$ is mutual capacitive coupling constant. 

Using commutation relations from Eq.~(\ref{Canonical commutation relation for bosonic fields}), the Hamiltonian (\ref{Hamiltonian of each quantum circuit}) and (\ref{Total Hamiltonian}), and the definition of the current operators
%
one can present the equations of motion for the currents and the charges  in a form of Langevin equations~\cite{ArturEquil,IdrisovThermalDecay}, namely 
\begin{equation}
\label{Langevin equations}
\begin{split}
& \partial_t Q_{\alpha}(t)=j_{\rm in,\alpha}(t)-j_{\rm out,\alpha}(t), \\
& j_{\rm out,\alpha}(t)=[Q_{\alpha}(t)+\lambda Q_{-\alpha}(t)]/\tau_{\rm c}+j_{\rm c,\alpha}(t), \\
& j_{\rm out,\alpha}(t)+j^{\rm c}_{\rm out,\alpha}(t)=j_{\rm in,\alpha}(t)+j_{\rm c,\alpha}(t),
\end{split}
\end{equation}
where $\lambda=C/C_{\rm ud} \le 1$ is the dimensionless coupling constant, $\tau_{\rm c}=R_{\rm q} C$ is the charge relaxation time, $R_{\rm q}=2\pi \hbar/e^2$ is the resistance quantum, and $-\alpha=\rm d,u$ simply denotes swapping the indexes $\alpha=\rm u,d$. Here, the first equation expresses the conservation of charge in the  circuit $\alpha$. The second line is the Langevin equation, which has the simple meaning: The outgoing current acquires one contributions $[Q_{\alpha}(t)+\lambda Q_{-\alpha}(t)]/\tau_{\rm c}$ from the time dependent potential of an Ohmic contact $\alpha$, and the second contribution $j_{\rm c,\alpha}$ that can be viewed as a Langevin source. The third equation expresses the conservation of incoming and outgoing currents at each Ohmic contact. 

The Eqs.~(\ref{Langevin equations}) can be easily solved, and the results may be presented as
\begin{equation}
\label{Scattering equation in frequency representation}		
\mathbf{J}_{\rm out}(\omega)=\mathcal{\hat U}(\omega) \mathbf{J}_{\rm in}(\omega),
\end{equation}
where the four component vectors of outgoing and incoming currents are given by 
$\mathbf{J}_{\rm out}=
(j_{\rm out,u}, j_{\rm out,d}, j^{\rm c}_{\rm out,u}, j^{\rm c}_{\rm out,d})$, and $\mathbf{J}_{\rm in}=(
j_{\rm in,u} , j_{\rm in,d} , j_{\rm c,u} , j_{\rm c,d})$
The scattering matrix in Eq.~(\ref{Scattering equation in frequency representation}) has the form 
\begin{equation}
\label{Scattering matrix}
\mathcal{\hat U}(\omega)=
\begin{bmatrix}
\mathcal{A}& \mathcal{B} & \mathcal{C}& -\mathcal{B} \\
\mathcal{B} & \mathcal{A} & -\mathcal{B} & \mathcal{C}\\
\mathcal{C} & -\mathcal{B} & \mathcal{A} & \mathcal{B} \\
-\mathcal{B} & \mathcal{C}& \mathcal{B} & \mathcal{A}
\end{bmatrix},
\end{equation}
where $\mathcal{A}(\omega)=[i\omega \tau_{\rm c}+\lambda^2-1]/[(\omega \tau_{\rm c}+i)^2+\lambda^2]$, $\mathcal{B}(\omega)=i\lambda \omega \tau_{\rm c}/[(\omega \tau_{\rm c}+i)^2+\lambda^2]$, and $\mathcal{C}(\omega)=\omega \tau_{\rm c} [\omega \tau_{\rm c}+i]/[(\omega \tau_{\rm c}+i)^2+\lambda^2]$. One can easily check the unitary of the  scattering matrix: $\mathcal{\hat U}^{\dagger}(\omega) =\mathcal{\hat U}(\omega)^{-1}$, which reflects the conservative character of the equations of motion, before Langevin sources are traced out. It is also worth mentioning that at $\lambda=0$ or $ 1$ one recovers the results for scattering matrix obtained in Refs.~\cite{ArturEquil,IdrisovThermalDecay,IdrisovDephasing}.

Next, we assume that currents originating from Ohmic contacts are equilibrium.  Thus, the  two-point correlation functions of the incoming currents as well as the Langevin sources are given by the equilibrium spectral function~\cite{Landau}
\begin{equation}
\label{FDT for currents}
\langle \delta j_{l, \alpha}(\omega) \delta j_{k, \beta}(\omega^{\prime})\rangle=2\pi \delta_{lk}\delta_{\alpha \beta} \delta(\omega+\omega^{\prime}) S_{l,\alpha}(\omega),
\end{equation}
where $S_{l,\alpha}(\omega)=R^{-1}_{\rm q} \omega/(1-e^{-\omega/T_{l, \alpha}})$ and $l,k={\rm in, c}$, and $\alpha,\beta=\rm u,d$. These correlation functions are used for further calculations. Note, however, the temperatures of Ohmic contacts $T_{\rm c,u/d}$ are yet unknown. They will be found below.

\textit{Heating effect and thermal drag current.} 
In a chiral QH channel, the energy current operator is equal to the energy density operator multiplied by group velocity ${\cal J}_{l,\alpha}(t)=(v^2/4\pi)[\partial_x \phi_{l,\alpha}(x,t)]^2$. Using the equation of motion for the bosonic field, $\phi_{l,\alpha}(x,t)$, one can rewrite the energy current as 
${\cal J}_{l,\alpha}(t)=(R_{\rm q}/2) j^2_{l,\alpha}(t),$
where $j_{l,\alpha}(t)$ are given in Eqs.~(\ref{Langevin equations}) and (\ref{Scattering equation in frequency representation}). The heat current can be obtained by subtracting the vacuum (zero temperature) contribution: 
\begin{equation}
\label{Definition of heat current operator}
J_{l,\alpha}=(R_{\rm q}/2) [\langle j^2_{l,\alpha}\rangle-\langle j^2_{l,\alpha}\rangle_{\rm vac}]. 
\end{equation}
For a ballistic equilibrium channel at the filling factor $\nu=1$ and temperature $T$ one uses the noise spectral function Eq.~(\ref{FDT for currents})  to arrive at the known result ~\cite{ArturEquil,Pekola}: $J= \pi k^2_B T^2/12 \hbar\equiv J_{\rm Q}$, where $J_Q$ is called the heat flux quantum.

Next, in order to find temperatures $T_{\rm c,u}$ and $T_{\rm c,d}$ in Eq.~(\ref{FDT for currents}), we solve self-consistently the energy balance equations, i.e.,  the conservation of incoming and outgoing heat currents at the Ohmic contacts (coupling to phonons, neglected in the present paper, can also be taken into account as in Ref.\ \cite{Sivre2018,Sivre2019,Rosenblatt}). These equations read $J_{\rm c,u}=J^{\rm c}_{\rm out,u}$, $J_{\rm c,d}=J^{\rm c}_{\rm out,d}$, where the heat currents are defined in Eq.~(\ref{Definition of heat current operator}).  Using the relations between incoming and outgoing currents  Eq.~(\ref{Scattering equation in frequency representation})  and spectral noise functions  from the Eq.~(\ref{FDT for currents}), one arrives at the system of coupled nonlinear equations for temperatures of Ohmic contacts: 
\begin{equation}
\label{System of equation to find temperature Tcu and Tcd with lambda}
\begin{split}
\frac{\pi T^2_{\rm c,u}}{12}= \frac{\pi T^2_{\rm in,u}}{12}+J^{\mathcal{B}}_{\substack{\rm in,d\\ \rm in,u}}(\lambda)+J^{\mathcal{A}}_{\substack{\rm c,u\\ \rm in,u}}(\lambda)+J^{\mathcal{B}}_{\substack{\rm c,d\\ \rm in,u}}(\lambda),\\
\frac{\pi T^2_{\rm c,d}}{12}=\frac{\pi T^2_{\rm in,d}}{12}+J^{\mathcal{B}}_{\substack{\rm in,u\\ \rm in,d}}(\lambda)+J^{\mathcal{A}}_{\substack{\rm c,d\\ \rm in,d}}(\lambda)+J^{\mathcal{B}}_{\substack{\rm c,u\\ \rm in,d}}(\lambda).
\end{split}
\end{equation}
The functions on the right hand side are given by 
\begin{equation}
\label{Main integral}	
\begin{split}
& J^{i}_{\substack{l,\alpha\\ k,\beta}}(\lambda)=\pi[T^2_{l,\alpha}\cdot I_{i}(\lambda, \tau_{\rm c} T_{l,\alpha})-T^2_{k,\beta}\cdot I_{i}(\lambda, \tau_{\rm c} T_{k,\beta})]/12, \\ 
& I_{i}(\lambda,a)=\frac{6}{(\pi a)^2}\int_0^{\infty}\!    \frac{dz z g_i(z,\lambda)}{\exp(z/a)-1}, \quad i=\mathcal{A},\mathcal{B},
\end{split}
\end{equation}
where $g_{\mathcal{A}}(z,\lambda)=|iz+\lambda^2-1|^2/|(z+i)^2+\lambda^2|^2$,  $g_{\mathcal{B}}(z,\lambda)=\lambda^2 z^2/|(z+i)^2+\lambda^2|^2$, and $z=\omega \tau_{\rm c} $ is the dimensionless integration variable. Note, that  the solution of Eq.~(\ref{System of equation to find temperature Tcu and Tcd with lambda}) must be an even function of $\lambda$, since $g_i(z,-\lambda)=g_i(z,\lambda)$.

We first concentrate on the limit of small temperatures,  ${\rm max}\{\tau_{\rm c} T_{l,\alpha}\} \ll 1$. In the case of ultimately  strong coupling, $\lambda=1$, the system of equations simplifies, and to the leading order one obtains equal temperatures:
\begin{equation}
\label{Tcu and Tcd temperatures in case of lambda equals to one and small TRC}
T^2_{\rm c,u}=T^2_{\rm c,d}=\left(T^2_{\rm in,u}+T^2_{\rm in,d} \right)/2.
\end{equation}
This result has a simple physical meaning: at $\lambda=1$ and small temperatures two Ohmic contacts merge into one, and the total incoming heat flux $\pi (T^2_{\rm in,u}+T^2_{\rm in,d})/12$  is  equally distributed between outgoing channels. In the case of weak coupling, $\lambda \ll 1$, the Eqs.\ (\ref{System of equation to find temperature Tcu and Tcd with lambda}) can be expanded to include corrections to the leading order in $\lambda$:
\begin{equation}
\label{Tcu and Tcd temperatures in case of lambda equals to zero and small TRC}
\begin{split}
& T^4_{\rm c,u}=T^4_{\rm in,u}+2\lambda^2 \left(T^4_{\rm in,d}-T^4_{\rm in,u}\right), \\
& T^4_{\rm c,d}=T^4_{\rm in,d}+2\lambda^2 \left(T^4_{\rm in,u}-T^4_{\rm in,d}\right).
\end{split}
\end{equation}

In the opposite limit of large temperatures, ${\rm min}\{\tau_{\rm c}T_{l,\alpha}\} \gg 1$, one arrives at the result that holds for arbitrary values of $\lambda$ in the interval from $0$ to $1$~\cite{Supp},
\begin{equation}
\label{Tcu and Tcd temperatures in case of lambda equals to zero and large TRC}
\begin{split}
& T^2_{\rm c,u}=T^2_{\rm in,u}+3\lambda^2(T_{\rm in,d}-T_{\rm in,u})/2\pi \tau_{\rm c}, \\
& T^2_{\rm c,d}=T^2_{\rm in,d}+3\lambda^2(T_{\rm in,u}-T_{\rm in,d})/2\pi \tau_{\rm c}.
\end{split}
\end{equation}
This is because at large temperatures the interactions are effectively weak for the arbitrary strength of coupling.

In order to investigate the thermal drag effect, we assume that the lower circuit is passive and cold, $T_{\rm in,d}=0$, and concentrate on the most interesting limit of low temperatures:  $\tau_{\rm c}T_{\rm in,u} \ll 1$. In the case of strong coupling, $\lambda \to 1$, the temperatures of Ohmic contacts are equal (see Eq.\ (\ref{Tcu and Tcd temperatures in case of lambda equals to one and small TRC})), therefore  the total incoming heat flux $J_{\rm in, u}=J_{\rm Q}$ in the upper incoming channel with the temperature $T_{\rm in, u}$ splits in two equal outgoing fluxes.  Thus the thermal drag current in the lower circuit takes its maximum value of 
\begin{equation}
\label{Heat curret at large lambda equals to one. Tind equals to zero and TRC small then one}
J_{\rm out, d}=J_{\rm in, u}/2.
\end{equation}  
For small $\lambda \ll 1$ the the thermal drag  current acquires the following form~\cite{Supp}
\begin{equation}
\label{Heat curret at large lambda equals to zero. Tind equals to zero and TRC small then one}
J_{\rm out, d}=(8\pi^2\lambda^2/5)( \tau_{\rm c} T_{\rm in,u})^2 J_{\rm in, u},
\end{equation}
i.e., it is suppressed by two small parameters, $\lambda$ and  $\tau_{\rm c} T_{\rm in,u}$. In what follows, we use Eqs.~(\ref{Heat curret at large lambda equals to one. Tind equals to zero and TRC small then one}) and (\ref{Heat curret at large lambda equals to zero. Tind equals to zero and TRC small then one})  to calculate the Fano factor.

\textit{Noise of thermal drag current.} 
The spectral density of heat current fluctuations at zero frequency and the Fano factor of the heat current are defined as
\begin{equation}
\label{Definition of noise}
\mathcal{S}_{l,\alpha}=\int dt \langle \delta J_{l, \alpha}(t) \delta J_{l, \alpha}(0)\rangle,\quad F_{l,\alpha}\equiv \mathcal{S}_{l,\alpha}/\langle J_{ l, \alpha}\rangle,
\end{equation}
where $\delta J_{l, \alpha}(t)=J_{l, \alpha}(t)-\langle J_{l, \alpha}(t) \rangle$. We first consider an equilibrium ballistic channel at the temperature $T$ as a reference. By substituting the operator of heat current  from Eq.~(\ref{Definition of heat current operator}) into Eq.~(\ref{Definition of noise}) and applying the Wick's theorem, we arrive at the following result~\cite{Pekola,Supp}:
\begin{equation}
\label{Zero frequency heat current noise. Result for filling factor one}	
\mathcal{S}=2k_B TJ_Q\equiv \mathcal{S}_{\rm Q}\quad {\rm and}\quad F=2k_B T\equiv F_Q.
\end{equation}Thus, the upper incoming channel carries the heat current noise $\mathcal{S}_{\rm in,u}=\mathcal{S}_{\rm Q}$ with the temperature $T_{\rm in, u}$, and the Fano factor is equal to $F_{\rm Q}$.

Next, using Eqs.~(\ref{Scattering equation in frequency representation}) and (\ref{Scattering matrix})  after some algebra we arrive at the following expression
\begin{equation}
\label{Zero frequency noise in down channel}	
\mathcal{S}_{\rm out, d}=\sum_{\substack{lk\\ \alpha \beta}} \int \frac{d\omega}{2\pi} S_{l,\alpha}(-\omega)\mathcal{M}_{l,\alpha; k,\beta}(-\omega,\omega)S_{k, \beta}(\omega), 
\end{equation}
where $S_{l,\alpha}(\omega)$ are given in Eq.~(\ref{FDT for currents}), and we introduced the function $\mathcal{M}_{l,\alpha; k,\beta}(-\omega,\omega)=m_{l,\alpha}(-\omega) m_{k,\beta}(\omega)/2$ with $m_{\rm in, u}(\omega)=|\mathcal{B}(\omega)|^2$, $m_{\rm in, d}(\omega)=|\mathcal{A}(\omega)|^2$, $m_{\rm c, d}(\omega)=|\mathcal{C}(\omega)|^2$, and $m_{\rm c, u}(\omega)=m_{\rm in,u}(\omega)$. This general result can be applied to a number of physical situations. Below, we concentrate on the noise of the thermal drag current in the most interesting limit of  $T_{\rm in,d}=0$ and $\tau_{\rm c}T_{\rm in,u} \ll 1$ and in the regimes of strong and weak coupling.

In the limit $\lambda =1$, one can set $m_{\rm in,u}=m_{\rm in,d}=m_{\rm c,u}=m_{\rm c,d} = 1/4$,  and the straightforward calculation gives~\cite{Supp}
\begin{equation}
\label{Zero frequency noise for down heat current at large lambda equals to one}
\mathcal{S}_{\rm out, d}=(3 \mathcal{I}/2\pi^2)\mathcal{S}_{\rm in, u},
\end{equation}
where  $\mathcal{I}\approx 2.5782$. By using the expression for heat current in the lower outgoing channel,  Eq.~(\ref{Heat curret at large lambda equals to one. Tind equals to zero and TRC small then one}), we obtain the Fano factor of the thermal drag current: 
\begin{equation}
\label{Fano factor in case of large lambda equals to one and Tind equals to zero. TinuRC smaller one}
F_{\rm out, d}/F_{\rm in, u} \approx 1.5673.
\end{equation}
In the limit of weak coupling, $\lambda \ll 1$, one obtains~\cite{Supp}
\begin{equation}
\label{Zero frequency noise for down heat current at small lambda equals to zero}
\mathcal{S}_{\rm out, d}=(3\lambda^2\mathcal{K}/2) ( \tau_{\rm c} T_{\rm in,u}) ^2\mathcal{S}_{\rm in, u},
\end{equation}
where $\mathcal{K}\approx 10 $.  By using Eq.~(\ref{Heat curret at large lambda equals to zero. Tind equals to zero and TRC small then one}) for the heat current, one obtains  the Fano factor: 
\begin{equation}
\label{Fano factor in case of large lambda equals to zero and Tind equals to zero. TinuRC smaller one}
F_{\rm out, d}/F_{\rm in, u} \approx 0.9498.
\end{equation}

The above result deserves additional discussion. First, we note that the Fano factor of the noise of  the equilibrium heat current can be estimated as a size of the typical fluctuation of heat, that is of the order of the average heat current times the correlation time. In equilibrium, the correlation time of the heat current noise is of the order of $1/T$ (the only available time scale). That is why the Fano factor of the equilibrium heat current noise scales linearly with the temperature (see Eq.\ (\ref{Zero frequency heat current noise. Result for filling factor one})). Therefore, it appears somewhat surprising, from the first glance, that in the weak coupling regime, $\lambda\ll 1$, the Fano factor of the thermal drag current is close to the one of the incoming equilibrium channel with the temperature $T_{\rm in,u}$, despite the fact that the effective temperature of the thermal  drag current scales as $\lambda \tau_c T^2_{in,u}$, i.e., it is smaller by the dimensionless factor $\lambda \tau_c T_{in,u}\ll 1$.
The explanation of this effect lies in the fact that in the weak coupling regime the thermal drag effect can be viewed as  essentially non-equilibrium rare Poissonian process of the emission and re-absorption of photons between the upper and  lower parts of the circuit. In this case the Fano factor acquires the values of the order of the average energy of transmitted photons, that is of the order of the temperature of the incoming equilibrium channel.

To summarize, we have proposed and theoretically analyzed a strongly interacting mesoscopic electron system for studying the thermal drag effect. The system that is based on QH edge states perfectly coupled to Ohmic contacts is accessible to modern experiments. It consists of an active circuit, where the heat is generated, and a passive circuit, where the heat flux is induced by non-local Coulomb interactions. The model of the system is exactly integrable with the help of non-equilibrium bosonization technique. We have calculated the thermal drag current, the corresponding zero frequency noise power, and the Fano factor of the thermal noise.  It has been  shown that depending on the interaction strength the Fano factor can be larger or smaller compared to the Fano factor of the equilibrium ballistic channel.      

EGI acknowledges financial support from the National Research Fund Luxembourg under 
Grants CORE 13579612. EVS acknowledges the financial support from the Swiss National Science Foundation.


\begin{thebibliography}{99}
	
\bibitem{Narozhny}
B. N. Narozhny and A. Levchenko, Rev. Mod. Phys. 88, 025003 (2016)
	
\bibitem{Levchenko}
A. Levchenko and A. Kamenev, Phys. Rev. Lett. 101, 216806 (2008)
	
\bibitem{Raichev}
O. E. Raichev, G. M. Gusev, F. G. G. Hernandez, A. D. Levin, and A. K. Bakarov
Phys. Rev. B 102, 195301 (2020)

\bibitem{Strait}
J. H. Strait, G. Holland, W. Zhu, C. Zhang, B. R. Ilic, A. Agrawal, D. Pacifici, and H. J. Lezec, Phys. Rev. Lett. 123, 053903 (2019)

\bibitem{Gurevich}
V. L. Gurevich and M. I. Muradov, Journal of Experimental and Theoretical Physics 121, 9981006 (2015) 

\bibitem{Joulain}
P. Ben-Abdallah, S. Biehs, and K. Joulain, Phys. Rev. Lett. 107, 114301 (2011)

\bibitem{Bhandari}
B. Bhandari, G. Chiriac\`{o}, P. A. Erdman, R. Fazio, and F. Taddei,
Phys. Rev. B 98, 035415 (2018)

\bibitem{Abdallah}
P. Ben-Abdallah, Phys. Rev. B 99, 201406(R) (2019)

\bibitem{Berdanier}
W. Berdanier, T. Scaffidi, and J. E. Moore, Phys. Rev. Lett. 123, 246603 (2019)
	
\bibitem{Venturelli}
D. Venturelli, R. Fazio, and V. Giovannetti, Phys. Rev. Lett. 110, 256801 (2013)

\bibitem{Zhang}
Y. Zhang, G. Lin, and J. Chen, Phys. Rev. E 91, 052118 (2015)

\bibitem{Koski}
J. V. Koski, V. F. Maisi, J. P. Pekola, and D. V. Averin,
PNAS 111, 13786 (2014)

\bibitem{Whitney}
R. S. Whitney, R. S\'{a}nchez, F. Haupt, and J. Splettstoesser,
Physica E 75, 257 (2016)

\bibitem{Ruokola}
T. Ruokola, and T. Ojanen, Phys. Rev. B 83, 241404 (2011).

\bibitem{Moskalets}
D. S\'{a}nchez and M. Moskalets, Entropy 22(9), 977 (2020)

\bibitem{Zubair}
Z. Iftikhar, {\it Charge Quantization and Kondo Quantum
Criticality in Few-Channel Mesoscopic Circuits}, 1st ed.
(Springer International Publishing, 2018)

\bibitem{Cavanna}
S. Jezouin, F. D. Parmentier, A. Anthore, U. Gennser,
A. Cavanna, Y. Jin, and F. Pierre, Science 342, 601
(2013)

\bibitem{Jezouin2016}
S. Jezouin, Z. Iftikhar, A. Anthore, F. D. Parmentier,
U. Gennser, A. Cavanna, A. Ouerghi, I. P. Levkivskyi,
E. Idrisov, E. V. Sukhorukov, L. I. Glazman,
and F. Pierre, Nature 536, 58 (2016).

\bibitem{IdrisovThermalDecay}
E. G. Idrisov, I. P. Levkivskyi, and E. V. Sukhorukov,
Phys. Rev. B 96, 155408 (2017).

\bibitem{Furusaki}
A. Furusaki and K. A. Matveev, Phys. Rev. B 52, 16676 (1995)

\bibitem{ArturEquil}
A. O. Slobodeniuk, I. P. Levkivskyi, and E. V. Sukhorukov, Phys. Rev. B 88, 165307 (2013).

\bibitem{Sivre2018}
E. Sivre, A. Anthore, F. D. Parmentier, A. Cavanna, U. Gennser, A. Ouerghi, Y. Jin, and F. Pierre, Nature Physics 14, 145 (2018).

\bibitem{Sivre2019}
C. Lin, M. Hashisaka, T. Akiho, K. Muraki, and T. Fujisawa,
Nature Communications 12, 131 (2021).

\bibitem{Clerk}
A. A. Clerk, P. W. Brouwer, and V. Ambegaokar, Phys.
Rev. Lett. 87, 186801 (2001).

\bibitem{IdrisovDephasing}
E. G. Idrisov, I. P. Levkivskyi, and E. V. Sukhorukov,
Phys. Rev. Lett. 121, 026802 (2018).

\bibitem{Pierre0}
H. Duprez, E. Sivre, A. Anthore, A. Aassime, A. Cavanna,
U. Gennser, and F. Pierre, Science 366, 1243
(2019).

\bibitem{Iftikhar2015}
Z. Iftikhar, S. Jezouin, A. Anthore, U. Gennser, F. D.
Parmentier, A. Cavanna, and F. Pierre, Nature 526, 233
(2015).

\bibitem{Iftikhar2018}
Z. Iftikhar, A. Anthore, A. K. Mitchell, F. D. Parmentier,
U. Gennser, A. Ouerghi, A. Cavanna, C. Mora, P. Simon,
and F. Pierre, Science 360, 1315 (2018).

\bibitem{Heiblum1}
M. Banerjee, M. Heiblum, A. Rosenblatt, Y. Oreg, D. E.
Feldman, A. Stern, and V. Umansky, Nature 545, 75
(2017).

\bibitem{Heiblum2}
M. Banerjee, M. Heiblum, V. Umansky, D. E. Feldman,
Y. Oreg, and A. Stern, Nature 559, 205 (2018).

\bibitem{Mross}
D. F. Mross, Y. Oreg, A. Stern, G. Margalit, and
M. Heiblum, Phys. Rev. Lett. 121, 026801 (2018).



\bibitem{Wen}
X. G. Wen, Phys. Rev. B 41, 12838 (1990).

\bibitem{Giamarchi}
T. Giamarchi, {\it Quantum Physics in One Dimension}
(Claverdon Press Oxford, 2004).

\bibitem{Landau}
E. M. Lifshitz and L. P. Pitaevskii, {\it Statistical Physics,
Part 2: (Landau and Lifshits Course of Theoretical
Physics, Vol.9)} (Butterworth-Heinemann Oxford, 1980).



\bibitem{Pekola}
J. P. Pekola and B. Karimi, Rev. Mod. Phys. 93, 041001 (2021)

\bibitem{Rosenblatt}
A. Rosenblatt, S. Konyzheva, F. Lafont, N. Schiller,
J. Park, K. Snizhko, M. Heiblum, Y. Oreg, and V. Umansky,
Phys. Rev. Lett. 125, 256803 (2020).

\bibitem{Supp}
See Supplemental Material at [] for the detailed
calculations of temperature, heat current and corresponding noise

\end{thebibliography}

\newpage
\begin{widetext}
\begin{center}
	{\large{\bf Thermal drag effect in quantum Hall circuits \\ Supplemental material}}\\
	\vspace{4mm}
\end{center}

\section{A: Heating effects}
\label{Sec:A}
In the case of $\lambda=1$, from Eq.~(11) of main text, we have  $g_{\mathcal{A}}(z,\lambda=1)=g_{\mathcal{A}}(z,\lambda=1)=1/(z^2+4)$ and the asymptotics of the integral in Eq.~(11) of main text have the simple form
\begin{equation}
	\label{Integral Ii for lambda=1}
	I_{i}(\lambda=1,a)=\frac{6}{(\pi a)^2}\int_0^{\infty}    \frac{dz z}{e^{z/a}-1} \frac{1}{z^2+4}\simeq 
	\begin{cases}
		1/4-(\pi a)^2/40, \quad a \ll 1, \\
		3/2\pi a, \quad a \gg 1.
	\end{cases}
\end{equation}
Substituting these asymptotics into Eq.~(10) from the main text, we further obtain the final results for the temperatures $T_{\rm c,u}$ and $T_{\rm c,d}$ in Eq.~(12) of the main text.

In the case of weak coupling, $\lambda \ll 1$, one can substitute the expansion of amplitudes of scattering matrix  $g_{\mathcal{A}}(z,\lambda \ll 1) \approx [1-4\lambda^2 z^2/(1+z^2)^2]/(1+z^2)$ and $g_{\mathcal{B}}(z,\lambda \ll 1) \approx \lambda^2 z^2/(1+z^2)^2$ into Eq.~(11) from main text and get 
\begin{equation}
	\label{Expansion of IA and IB in lambda less then one}
	I_{\mathcal{A}}(\lambda \ll 1, a) \simeq I_{\mathcal{A}}(\lambda=0,a)-4\lambda^2 \cdot \delta I_{\mathcal{A}}(\lambda=0,a), \quad \text{and} \quad I_{\mathcal{B}}(\lambda \ll 1, a)\simeq \lambda^2 \cdot \delta I_{\mathcal{B}}(\lambda=0, a),
\end{equation}
where the asymptotics of integrals on the right hand side are given by 
\begin{equation}
	\label{IA(lambda equals zero)}
	I_{\mathcal{A}}(\lambda=0, a) =\frac{6}{(\pi a)^2} \int_0^{\infty} \frac{dz z}{e^{z/a}-1} \frac{1}{1+z^2} \simeq 
	\begin{cases}
		1-2(\pi a)^2/5, \quad a \ll 1, \\
		3/\pi a, \quad a \gg 1,
	\end{cases}
\end{equation}
\begin{equation}
	\label{deltaIA(lambda equals zero)}	
	\delta I_{\mathcal{A}}(\lambda=0, a) = \frac{6}{(\pi a)^2} \int_0^{\infty}
	\frac{dz z}{e^{z/a}-1} \frac{z^2}{(1+z^2)^3} \simeq
	\begin{cases}
		2(\pi a)^2/5, \quad a \ll 1, \\
		3/2\pi a, \quad a \gg 1,
	\end{cases}	
\end{equation}
\begin{equation}
	\label{deltaIB(lambda equals zero)}	
	\delta I_{\mathcal{B}}(\lambda=0, a) =\frac{6}{(\pi a)^2} \int_0^{\infty}
	\frac{dz z}{e^{z/a}-1} \frac{z^2}{(1+z^2)^2} \simeq
	\begin{cases}
		2(\pi a)^2/5, \quad a \ll 1, \\
		3/8\pi a, \quad a \gg 1.
	\end{cases}	
\end{equation}

Substituting the above asymptotics in the case of ${\rm max}\{\tau_{\rm c}T_{l,\alpha}\} \ll 1$ ($a \ll 1$) into Eq.~(10) from the main text, one obtains the following system of equations
\begin{equation}
	\label{System of equations to find Tcu Tcd for TRC less one}	
	\begin{cases}	
		\begin{split}
			& \frac{\pi T^2_{\rm c,u}(\lambda)}{12}-\frac{\pi T^2_{\rm in,u}}{12}=\lambda^2 \left[\frac{\pi T^2_{\rm in,d}}{12} \frac{2(\pi \tau_{\rm c}T_{\rm in,d})^2}{5}-\frac{\pi T^2_{\rm in,u}}{12} \frac{2(\pi \tau_{\rm c} T_{\rm in,u})^2}{5}\right]\\
			& +\frac{\pi T^2_{\rm c,u}(\lambda)}{12} \left[1-\frac{2(\pi \tau_{\rm c}T_{\rm c,u}(\lambda))^2}{5}-4 \lambda^2 \frac{2(\pi \tau_{\rm c} T_{\rm c,u}(\lambda))^2}{5} \right]\\
			& -\frac{\pi T^2_{\rm in,u}}{12} \left[1-\frac{2(\pi \tau_{\rm c}T_{\rm in,u})^2}{5}-4 \lambda^2 \frac{2(\pi \tau_{\rm c} T_{\rm in,u})^2}{5} \right]\\
			& +\lambda^2 \left[\frac{\pi T^2_{\rm c,d}(\lambda)}{12} \frac{2(\pi \tau_{\rm c}T_{\rm c,d}(\lambda))^2}{5}-\frac{\pi T^2_{\rm in,u}}{12} \frac{2(\pi \tau_{\rm c} T_{\rm in,u})^2}{5}\right],
		\end{split} \\
		& \\
		\begin{split}
			& \frac{\pi T^2_{\rm c,d}(\lambda)}{12}-\frac{\pi T^2_{\rm in,d}}{12}=\lambda^2 \left[\frac{\pi T^2_{\rm in,u}}{12} \frac{2(\pi \tau_{\rm c}T_{\rm in,u})^2}{5}-\frac{\pi T^2_{\rm in,d}}{12} \frac{2(\pi \tau_{\rm c} T_{\rm in,d})^2}{5}\right]\\
			& +\frac{\pi T^2_{\rm c,d}(\lambda)}{12} \left[1-\frac{2(\pi \tau_{\rm c}T_{\rm c,d}(\lambda))^2}{5}-4 \lambda^2 \frac{2(\pi \tau_{\rm c} T_{\rm c,d}(\lambda))^2}{5} \right]\\
			& -\frac{\pi T^2_{\rm in,d}}{12} \left[1-\frac{2(\pi \tau_{\rm c}T_{\rm in,d})^2}{5}-4 \lambda^2 \frac{2(\pi \tau_{\rm c} T_{\rm in,d})^2}{5} \right]\\
			& +\lambda^2 \left[\frac{\pi T^2_{\rm c,u}(\lambda)}{12} \frac{2(\pi \tau_{\rm c}T_{\rm c,u}(\lambda))^2}{5}-\frac{\pi T^2_{\rm in,d}}{12} \frac{2(\pi \tau_{\rm c} T_{\rm in,d})^2}{5}\right],
		\end{split} 
	\end{cases}
\end{equation}
The solution of these equations for ${\rm max}\{\tau_{\rm c}T_{l,\alpha}\} \ll 1$ is given by Eq.~(13) from the main text.

In the opposite regime, ${\rm min}\{\tau_{\rm c}T_{l,\alpha}\} \gg 1$, one can obtain the result which is non-perturbative in $\lambda$. Namely, by calculating the integrals in this regime ($a \gg 1$) we get the following asymptotic results 
\begin{equation}
	\label{Asymptotics IA. Nonperturbative lambda}	
	\begin{split}	
		& I_{\mathcal{A}}(\lambda,a)= a \int_0^{\infty} dz \frac{z/a}{e^{z/a}-1} \frac{z^2+(\lambda^2-1)^2}{z^4+2(\lambda^2+1)z^2+(\lambda^2-1)^2}=[a \to \infty] \\
		& \simeq a \int_0^{\infty} dz \frac{z^2+(\lambda^2-1)^2}{z^4+2(\lambda^2+1)z^2+(\lambda^2-1)^2}=a \cdot \mathcal{F}_{\mathcal{A}}(\lambda), \quad \mathcal{F}_{\mathcal{A}}(\lambda)=
		\frac{\pi}{4}(2-\lambda^2),
	\end{split}
\end{equation}
and similarly 
\begin{equation}
	\label{Asymptotics IB. Nonperturbative lambda}
	\begin{split}	
		& I_{\mathcal{B}}(\lambda,a)= a \int_0^{\infty} dz \frac{z/a}{e^{z/a}-1} \frac{z^2 \lambda^2}{z^4+2(\lambda^2+1)z^2+(\lambda^2-1)^2}=[a \to \infty] \\
		& \simeq  a \int_0^{\infty} dz \frac{z^2\lambda^2}{z^4+2(\lambda^2+1)z^2+(\lambda^2-1)^2}=a \cdot \mathcal{F}_{\mathcal{B}}(\lambda), \quad \mathcal{F}_{\mathcal{B}}(\lambda)=\frac{\pi \lambda^2}{4}.
	\end{split}
\end{equation}
After the substitution of Eq.~(\ref{Asymptotics IA. Nonperturbative lambda}) and (\ref{Asymptotics IB. Nonperturbative lambda}) into Eq.~(10) from the  main text, the system of equations to find the temperatures of Ohmic contacts takes the form
\begin{equation}
	\label{System of equation nonperturbative result in lambda but TRC larger then 1}
	\begin{cases}
		\frac{\pi T^2_{\rm c,u}(\lambda)}{12}=\frac{\pi T^2_{\rm in,u}}{12}+\left[\frac{\pi T^2_{\rm in,d}}{12} \frac{6}{\pi \tau_{\rm c} T_{\rm in,d}}-\frac{\pi T^2_{\rm in,u}}{12} \frac{6}{\pi \tau_{\rm c} T_{\rm in,u}}\right]\cdot \mathcal{F}_{\mathcal{B}}(\lambda)\\
		+\left[\frac{\pi T^2_{\rm c,u}}{12} \frac{6}{\pi \tau_{\rm c} T_{\rm c, u}}-\frac{\pi T^2_{\rm in,u}}{12} \frac{6}{\pi \tau_{\rm c}T_{\rm in,u}}\right]\cdot \mathcal{F}_{\mathcal{A}}(\lambda)+\left[\frac{\pi T^2_{\rm c,d}}{12} \frac{6}{\pi \tau_{\rm c} T_{\rm c,d}}-\frac{\pi T^2_{\rm in,u}}{12} \frac{6}{\pi \tau_{\rm c} T_{\rm in,u}}\right]\cdot \mathcal{F}_{\mathcal{B}}(\lambda), \\
		\\
		\frac{\pi T^2_{\rm c,d}(\lambda)}{12}=\frac{\pi T^2_{\rm in,d}}{12}+\left[\frac{\pi T^2_{\rm in,u}}{12} \frac{6}{\pi \tau_{\rm c} T_{\rm in,u}}-\frac{\pi T^2_{\rm in,d}}{12} \frac{6}{\pi \tau_{\rm c} T_{\rm in,d}}\right]\cdot \mathcal{F}_{\mathcal{B}}(\lambda)\\
		+\left[\frac{\pi T^2_{\rm c,d}}{12} \frac{6}{\pi \tau_{\rm c} T_{\rm c, d}}-\frac{\pi T^2_{\rm in,d}}{12} \frac{6}{\pi \tau_{\rm c}T_{\rm in,d}}\right]\cdot \mathcal{F}_{\mathcal{A}}(\lambda)+\left[\frac{\pi T^2_{\rm c,u}}{12} \frac{6}{\pi \tau_{\rm c} T_{\rm c,u}}-\frac{\pi T^2_{\rm in,d}}{12} \frac{6}{\pi \tau_{\rm c} T_{\rm in,d}}\right]\cdot \mathcal{F}_{\mathcal{B}}(\lambda).
	\end{cases}	
\end{equation}
As a first approximation, we choose the solution at $\lambda=0$, namely, we substitute $T_{\rm c,u}=T_{\rm in,u}$ and $T_{\rm c,d}=T_{\rm in,d}$ into the right hand side. Therefore, only the first term on the right hand side survives, and we get the final solution at ${\rm min}\{\tau_{\rm c}T_{l,\alpha}\}\gg 1$
\begin{equation}
	\label{Nonperturbative in lambda result fro TRCggOne }
	\begin{split}
		& \frac{\pi T^2_{\rm c,u}(\lambda)}{12}=\frac{\pi T^2_{\rm in,u}}{12}+\frac{3 \lambda^2}{2 }\left[\frac{\pi T^2_{\rm in,d}}{12}\frac{1}{\pi \tau_{\rm c}T_{\rm in,d}}-\frac{\pi T^2_{\rm in,u}}{12}\frac{1}{\pi \tau_{\rm c} T_{\rm in,u}}\right], \\
		& \frac{\pi T^2_{\rm c,d}(\lambda)}{12}=\frac{\pi T^2_{\rm in,d}}{12}+\frac{3 \lambda^2}{2}\left[\frac{\pi T^2_{\rm in,u}}{12}\frac{1}{\pi \tau_{\rm c}T_{\rm in,u}}-\frac{\pi T^2_{\rm in,d}}{12}\frac{1}{\pi \tau_{\rm c} T_{\rm in,d}}\right].
	\end{split}
\end{equation}
This result is provided in Eq.~(14) of main text.

\section{B: Thermal drag current}
\label{Sec:B}
In this section we provide the calculation of outgoing heat current in down passive circuit in case of $\lambda \ll 1$ and $T_{\rm in,d}=0$ and $\tau_{\rm c} T_{\rm in,u} \ll 1$. Using the definition of heat current Eq.~(9) in main text one has 
\begin{equation}
	\label{General expression for heat current}
	\begin{split}
		& J_{\rm d}=\frac{R_{\rm q}}{2} \int \frac{d\omega}{2\pi}\left[|\mathcal{B(\omega)}|^2 (S_{\rm in,u}(\omega)-S_0(\omega))+|\mathcal{A(\omega)}|^2 (S_{\rm in,d}(\omega)-S_0(\omega))\right] \\
		& + \frac{R_{\rm q}}{2} \int \frac{d\omega}{2\pi}\left[|\mathcal{B(\omega)}|^2 (S_{\rm c,u}(\omega)-S_0(\omega))+|\mathcal{C(\omega)}|^2 (S_{\rm c,d}(\omega)-S_0(\omega))\right],
	\end{split}
\end{equation} 
where $S_0(\omega)=R^{-1}_{\rm q} \omega \theta(\omega)$ ground state contribution, $S_{\rm in,u}(\omega)=R^{-1}_{\rm q}\omega/(1-e^{-\omega/T_{\rm in,u}})$, $S_{\rm in, d}=S_0(\omega)$, $S_{\rm c,u}(\omega)=R^{-1}_{\rm q}\omega/(1-e^{-\omega/f_{\rm c,u}(\lambda) T_{\rm in,u}})$, $S_{\rm c,d}(\omega)=R^{-1}_{\rm q}\omega/(1-e^{-\omega/f_{\rm c,d}(\lambda) T_{\rm in,u}})$ and according to Eq.~(13) from main text $f_{\rm c,u}(\lambda)=\sqrt[4]{1+3\lambda^2}$, $f_{\rm c,d}(\lambda)=\sqrt[4]{2\lambda^2}$. Introducing the dimensionless integration variable $y=\omega/T_{\rm in,u}$, we write the heat current as a sum of four terms
\begin{equation}
	J_{\rm }=J_1+J_2+J_3+J_4,
\end{equation}   
where in the leading order with respect to $\tau_{\rm c} T_{\rm in,u} \ll 1$ they read
\begin{equation}
	\begin{split}	
		& J_1=\frac{T^2_{\rm in,u}}{2} \int \frac{dy}{2\pi} \frac{(\lambda \tau_{\rm c}T_{\rm in,u})^2 y^2}{\tau^4_{\rm c}T^4_{\rm in,u}y^4+2\tau^2_{\rm c}T^2_{\rm in,u} y^2 (1+\lambda^2)+(1-\lambda^2)^2}\left[\frac{y}{1-e^{-y}}-y\theta(y)\right] \\
		& \simeq \frac{T^2_{\rm in,u}}{2}(\lambda \tau_{\rm c} T_{\rm in,u})^2 \int \frac{dy}{2\pi} y^3 \left[\frac{1}{1-e^{-y}}-\theta(y)\right], 
	\end{split}
\end{equation}
\begin{equation}
	J_2=0,
\end{equation}
\begin{equation}
	\begin{split}	
		& J_3=\frac{T^2_{\rm in,u}}{2} \int \frac{dy}{2\pi} \frac{(\lambda \tau_{\rm c} T_{\rm in,u})^2 y^2}{\tau^4_{\rm c}T^4_{in,u}y^4+2\tau^2_{\rm c} T^2_{\rm in,u} y^2(1+\lambda^2)+(1-\lambda^2)^2}\left[\frac{y}{1-e^{-y/f_{\rm c,u}(\lambda)}}-y\theta(y)\right] \\
		& =(f_{\rm c,u}(\lambda) \to 1) \simeq \frac{T^2_{\rm in,u}}{2} (\lambda \tau_{\rm c} T_{\rm in,u})^2 \int \frac{dy}{2\pi} y^3 \left[\frac{1}{1-e^{-y}}-\theta(y)\right], 
	\end{split}
\end{equation}
\begin{equation}
	\begin{split}	
		& J_4=\frac{T^2_{\rm in,u}}{2} \int \frac{dy}{2\pi} \frac{\left[1+(\tau_{\rm c}T_{\rm in,u}y)^2\right](\tau_{\rm c} T_{\rm in,u}y)^2}{\tau^4_{\rm c}T^4_{\rm in,u}y^4+2\tau^2_{\rm c} T^2_{\rm in,u} y^2 (1+\lambda^2)+(1-\lambda^2)^2}\left[\frac{y}{1-e^{-y/f_{\rm c,d}(\lambda)}}-y\theta(y)\right] \\
		& \simeq \frac{T^2_{\rm in,u}}{2}(\tau_{\rm c}T_{\rm in,u})^2 \left[f_{\rm c,d}(\lambda)\right]^4\int \frac{dy}{2\pi} y^3 \left[\frac{1}{1-e^{-y}}-\theta(y)\right] =(\left[f_{\rm c,d}(\lambda)\right]^4 \simeq 2\lambda^2) \\
		& \simeq T^2_{\rm in,u}(\lambda \tau_{\rm c}T_{\rm in,u})^2 \int \frac{dy}{2\pi} y^3 \left[\frac{1}{1-e^{-y}}-\theta(y)\right].
	\end{split}
\end{equation}
Thus the sum of four terms gives the result presented in Eq.~(16) of the main text 
\begin{equation}
	J_{\rm d}=\frac{T^2_{\rm in,u}}{\pi} (\lambda \tau_{\rm c} T_{\rm in,u})^2\underbrace{\int dy y^3\left[\frac{1}{1-e^{-y}}-\theta(y)\right]}_{2\pi^4/15}=\frac{8\lambda^2 (\pi \tau_{\rm c} T_{\rm in,u})^2}{5}J_{\rm Q},
\end{equation}
where $J_{\rm Q}=\pi T^2_{\rm in,u}/12$.

\section{C: Noise of thermal current for ballistic channel}
\label{Sec:C}
Using the definition (9) of the heat flux in the main text, we get the following expression
for the noise power of thermal current
\begin{equation}
	\mathcal{S}_{\rm Q}(\omega)=(R_{\rm q}/2)^2\int dt e^{i\omega t}\left[\langle j^2(t)j^2(0)\rangle-\langle j^2(t)\rangle \langle j^2(0)\rangle \right].
\end{equation}
According to Wick's theorem $\langle j^2(t)j^2(0)\rangle=\langle j^2(t)\rangle \langle j^2(0)\rangle+2\langle j(t)j(0)\rangle \langle j(t)j(0)\rangle$, since each current operator is the combination of bosonic creation and annihilation operators. Subtracting the second term we get
\begin{equation}
	\mathcal{S}_{\rm Q}(\omega)=(R^2_{\rm q}/2)\int dt e^{i\omega t} \langle j(t)j(0)\rangle^2.
\end{equation}
Next, using the FDT in Eq.~(8) of the main text, we calculate the above expression explicitly 
\begin{equation}
	\begin{split}	
		& \mathcal{S}_{\rm Q}(\omega)=(R^2_{\rm q}/2) \prod^4_{i=1}\int \frac{d\omega_i}{(2\pi)^4} \int dt e^{i(\omega-\omega_1-\omega_3)t} \langle j(\omega_1)j(\omega_2)\rangle \langle j(\omega_3)j(\omega_4)\rangle=\frac{1}{2} \int \frac{d\omega_1}{2\pi} R_{\rm q} S(\omega-\omega_1) \cdot R_{\rm q} S(\omega_1) \\
		& =\frac{1}{2} \int \frac{d\omega_1}{2\pi} \frac{\omega-\omega_1}{1-e^{-(\omega-\omega_1)/T}} \cdot \frac{\omega_1}{1-e^{-\omega_1/T}}=\frac{T^3}{2} \int \frac{dy}{2\pi} \frac{\frac{\omega}{T}-y}{1-e^{-\frac{\omega}{T}+y}} \cdot \frac{y}{1-e^{-y}}. 
	\end{split}
\end{equation}
By introducing the dimensionless variable $x=y-\omega/2T$ and performing the integral with respect to $x$ we obtain
\begin{equation}
	\label{Finite frequency noise for ballistic channel}	
	\mathcal{S}_{\rm Q}(\omega)=\frac{T^3}{4\pi} \int dx \frac{\frac{\omega}{2T}-x}{1-e^{x-\frac{\omega}{2T}}} \cdot \frac{x+\frac{\omega}{2T}}{1-e^{-\left(x+\frac{\omega}{2T}\right)}}=\frac{\omega}{48\pi}[(2\pi T)^2+\omega^2][1+\coth(\omega/2T)].
\end{equation}
It is worth mentioning that the non-symmetrized noise is given by $\mathcal{S}_{\rm Q}(\omega)/2+\mathcal{S}_{\rm Q}(-\omega)/2$. Because of the quadratic term $\omega^2$ in square brackets the finite frequency noise at zero temperature $T=0$ does not vanish: $\mathcal{S}_{\rm Q}(\omega)=\omega^3 \text{sgn}(\omega)/24 \pi $. Finally, taking the limit $\omega \to 0$ in Eq.~(\ref{Finite frequency noise for ballistic channel}) we arrive to Eq.~(18) from the main text.

\section{D: Noise of thermal drag current}
Similar calculations as in Sec.~B can be done for the noise of heat current in the case of $\lambda \to 1$ and $\lambda \ll 1$ at $T_{\rm in,d}=0$ and $\tau_{\rm c} T_{\rm in,u} \ll 1$. In the case of $\lambda \to 1$ the result is given by Eq.~(20) from the main text, where 
\begin{equation}
	\label{Dimensionless integral 1}
	\mathcal{I}= \int \frac{dy}{16} P(y)P(-y) \approx 2.5782,
\end{equation}
where $P(y)=y/(1-e^{-y})+y\theta(y)+2y/(1-e^{-\sqrt{2}y})$ and $\theta(y)$ is the Heaviside step function. $\sqrt{2}$ in the integrand of $\mathcal{I}$ originates from Ohmic contact temperatures $T_{\rm c,u}=T_{\rm c,d}=T_{\rm in, u}/\sqrt{2}$. In the case of $\lambda \ll 1$, the heat currents noise in the passive circuit (see Eq.~(19) in the main text) contains sixteen terms. After some algebra similar to that of Sec.~B, one can show that the leading order behavior of the  noise of heat current in the lower channel for $\lambda \ll 1$ and $\tau_{\rm c} T_{\rm in,d} \ll 1$  is given by Eq.~(22) in the main text. 

\end{widetext}

\end{document}